\documentclass{aastex61}

\def\lesssim{\mathrel{\hbox{\rlap{\hbox{\lower4pt\hbox{$\sim$}}}\hbox{$<$}}}}
\def\gtrsim{\mathrel{\hbox{\rlap{\hbox{\lower4pt\hbox{$\sim$}}}\hbox{$>$}}}}

\def\agt{\gtrsim}
\def\alt{\lesssim}

\newcommand{\red}[1]{}

\newcommand{\be}{\begin{equation}}
\newcommand{\ee}{\end{equation}}
\newcommand{\bea}{\begin{eqnarray}}
\newcommand{\eea}{\end{eqnarray}}
\newcommand{\bdm}{\begin{displaymath}}
\newcommand{\edm}{\end{displaymath}}

\begin{document}
\title{Using eigenmode-mixing to measure or constrain the Sun's interior B-field }
\author{Curt Cutler}
\affiliation{Jet Propulsion Lab, 4800 Oak Grove Dr., Pasadena, CA 91109}
\affiliation{Theoretical Astrophysics, California Institute of Technology, Pasadena, California 91125}

\begin{abstract}
Understanding the generation and distribution of the Sun's interior magnetic (B-) field is a longstanding challenge.   Here we  describe how measurements of the Sun's oscillation eigenfunctions might be used to measure the Sun's interior B-field.  The B-field induces mode-mode couplings, causing
the angular patterns of the eigenfunctions to differ from simple $Y_{lm}$'s.  We concentrate on the magnetic coupling between modes with  
the same $(n,l)$ values and different but nearby $m$-values, since these non-axisymmetric couplings clearly cannot be due to the Sun's axisymmetric rotation and since for these cases, mode mixing is enhanced by the near-degeneracy of the mode frequencies.   We analyze magnetically-induced mode mixing in two stages of increasing complexity:  first neglecting mode damping, and then incorporating realistic damping rates.  We introduce a novel detection statistic that tests for the presence of non-axisymmetric mode-mixing in Solar Doppler data.  We show that our detection statistic is naturally robust against spatial aliasing.
We estimate our statistic's signal-to-noise ratio (SNR) as a function of the mode-mixing amplitude.  While B-induced mode-mixing is probably not detectable in a single mode pair,  we argue that the phase of the B-induced mixing should be approximately the same across a wide range of modes.  The total SNR then grows roughly as $N_{pair}^{1/2}$, where $N_{pair}$ is the number of mode pairs.  We conclude that B-induced mode-mixing  should be detectable for a fairly wide range of B-field magnitudes and geometries.   
 %
%
%
\end{abstract}
\keywords{
Sun: helioseismology -- oscillations --magnetic fields
}

\section{Introduction}
\label{Sec:Int}

Understanding the generation and distribution of the Sun's interior magnetic field (B-field) 
is a major and longstanding goal of solar physics.
%
In standard dynamo models, large-scale fluid flows
--especially differential rotation in the shear layer 
near the bottom
of the convection zone and the meridional flow from
equator to poles  --play central roles in generating and transporting magnetic field; see, e.g.~\citet{miesch05}.
Additionally, it has long been hypothesized that the Sun's core contains
a very strong toroidal B-field--possibly as large as $\sim 10^6 G$--which is left over from
the Sun's formation, and which has persisted because of the extremely high conductivity  in the Sun's core; see, e.g.,~\citet{gough_taylor84}.
%
%
%

Much of our knowledge of the Sun's interior comes from helioseismology.  Analysis methods for helioseismology data generally fall into one of two broad classes:  local helioseismology and  global helioseismology.  The method proposed in this paper falls within the category of  global helioseismology.  
Most (though certainly not all) work in global helioseismology
has focused on the information encoded in the Sun's oscillation  frequencies.  
However, 
in this paper we focus on information about the Sun's interior B-field that might be gleaned from measuring the mode shapes; 
i.e, the eigenfunctions as opposed to the eigenfrequencies.
%

 In this paper we concentrate on couplings between modes that have the same $(n,l)$ values and $m-$values that are separated by a small (non-zero) integer.
There are several motivations for this:  a) sunspots tend to aggregate at preferred, evenly spaced longitudes (between one and four longitudes depending on the particular epoch and the northern versus southern hemispheres), which suggests the presence of substantial low-$m$ power in the B-field distribution, b) modes with the same
$(n,l)$ values and neighboring $m$-values are nearly degenerate in frequency, and near-degeneracies amplify mode-mixing, and c)
couplings between modes with different $m$-values are clearly {\it not} 
due to axisymmetric flows.

We analyze magnetically-induced mode mixing in two stages of increasing complexity.  In the first stage, as a warm-up problem, we neglect the effects of mode damping.  In the second stage, we account for realistic levels of mode damping.   Our analysis motivates the introduction of a new, mode-mixing detection statistic that tests for the presence of non-axisymmetric mode-mixing in solar Doppler data.   A nice feature of our detection statistic is that it is quite insensitive 
to the effects of spatial aliasing. 

We estimate our statistic's expected signal-to-noise ratio, as a function of the amplitude of the mode-mixing (which in turn depends on the amplitude and distribution of the Sun's interior B-field).  While we find that B-induced Solar mode mixing is likely too weak to be measurable with a single mode pair, we argue that combining results from a large number of mode pairs should dramatically increase the power of the method.  The basic reason is that mode-mixing has a phase as well as an amplitude, so calculating our detection statistic produces a complex number.  As we argue below, it seems reasonable to assume
that over large regions in the space of mode-pairs, the "signal contribution" to the statistic should produce nearly the same phase.  So if one simply adds up these complex values for a large number of similar modes, the 
SNR for the mode-summed version of our detection statistic should grow roughly as $N_{pair}^{1/2}$, where $N_{pair}$ is the number of relevant mode pairs.  
With the mode-summed version of our detection statistic, we find that B-induced mode mixing should be measurable for a
fairly wide range of plausible B-field magnitudes and geometries.  

Of course, many of the ideas and methods in this paper already exist in the substantial literature on helioseismology.  E.g., a formalism for using eigenmode-mixings to measure the Sun's meridional flow, which emphasizes the importance of near-degeneracies, was developed and applied to SOHO data in a series of papers 
by~\cite{schad1,schad2,schad3,schad4}.   
Also, inference from eigenfunction mixing is clearly closely related to inference from the analysis of cross-spectra, and the latter has been used to investigate meridional and zonal flows in a large number of papers 
by Woodard, Schou and others; see, e.g.,~\cite{woodard12,woodard13}.
And of course there has been a good deal of other work exploring what helioseismology can say about the Sun's interior B-field.  E.g., 
\cite{gough1990} considered the effects of an axisymmetric magnetic field that is misaligned from the Sun's spin axis, 
while~\cite{antia2013} began the project of inferring the B-field in the convection zone from helioseismologically measured variations in the Sun's local angular velocity. 
What is novel in our paper, we believe, is  i) our emphasis on probing a truly non-axisymmetric B-field (not just a mis-aligned one) using measurements of eigenfunction-mixing between modes with different $m$ values, ii) our analysis of how results from nearly-degenerate perturbation theory get modified when 
realistic damping times for Solar modes are taken into account, and iii) our mode-mixing detection statistic, which we designed specifically to take realistic damping into account and to minimize the influence of spatial aliasing.

Finally, we mention that based on recent, asteroseismological measurements by the Kepler satellite~\cite{fuller_etal_2015} suggest that $\sim 20 \% $ of red giants in the Kepler sample have very strong core B-fields: $B \agt 10^5\,$G  and up to $\sim 10^7\,$G.  However the relevance of these new discoveries to our own Sun is questionable,  since at present the strong core B-fields are limited to red giants whose main sequence progenitors would have had masses in the range $1.6-2.0 M_{\odot}$, and therefore would have had convective cores where the B-fields could have been generated by dynamo action; 
see~\cite{stello_etal_2016}.
%

The rest of this paper is organized as follows.  In Sec.~\ref{Sec:backgd} we summarize some necessary background material, partly to establish notation.  In Sec.~\ref{Sec:mixing} we analyze the mixing of Solar oscillation modes with different $m$ values, resulting from small, non-axisymmetric perturbations.
As a warm-up problem, in Sec.~\ref{Sec:zero-damp} we first examine this mixing in the limit of zero mode damping.   In 
Sec.~\ref{Sec:mixing_with_damping} we analyze Solar mode mixing for realistic mode damping rates and  introduce our new mode-mixing detection statistic.
In Sec.~\ref{Sec:mixing_with_damping} we also estimate the signal-to-noise ratio (SNR) for this statistic, applied to a single mode pair, as a function of the strength of the B-field's coupling between those modes.  We find that mode-mixing is likely {\it not} observable for a single mode pair.  However, we argue in Sec.~\ref{sec:sum} that combining the results from a large number of mode pairs could yield a detectable $SNR$ for an interesting range of B-field distributions.  And while it might naively be supposed that spatial aliasing 
could significantly degrade our statistic's detection power, in Sec.~\ref{sec:aliasing} we show why this is {\it not} the case, both analytically and via a (simplified) numerical simulation. In Sec.~\ref{Sec:summary}  we summarize our results, call attention to a couple caveats, and very briefly discuss next steps.
\section{Background material}
\label{Sec:backgd}
\subsection{The signal and the noise}
\label{Sec:sig}
Our sign convention for modes is that a mode with frequency $\omega$ and azimuthal number $m$ has dependence $e^{i \omega t - i m \phi}$ (with the Sun rotating in the positive-$\phi$ direction).    Then to see the effect of rotation on mode frequencies, consider a mode that  in the Sun's co-rotating frame has the form   $e^{i \omega_{co} t - i m\tilde \phi}$, where $\tilde \phi \equiv \phi - \Omega t$ is the co-rotating azimuthal coordinate.   The main effect of rotation on mode shapes and frequencies is a kinematic one: the mode gets dragged forward by the Sun's rotation.  Then in the inertial frame the mode has (approximately) the form $e^{i \omega_{co} t - i m (\phi - \Omega t)}$, so the mode frequency in the inertial frame becomes $\omega \approx \omega_{co} + m \Omega$.
  
We will focus on the information encoded in the Doppler signal, $\delta \nu/\nu = (\vec v/c) \cdot \hat n$, where $\nu$ is the center of some
absorption line, $\vec v$ is the velocity of the fluid at the Solar surface (more correctly, at the photosphere for the particular line of interest), and 
$\hat n$ is the line of sight to the detector.  We denote this Doppler signal as $y = y(\theta,\phi,t)$, where
$(\theta, \phi)$ are (inertial-frame) spherical coordinates on the Sun's surface.
The (time-dependent part of) the fluid velocity $\vec v$  at the Sun's surface (radius $R$) can be decomposed as a sum of modes:
\be\label{vr}
\vec v(\theta,\phi,t) = \sum_{nlm} i\, \omega_{nlm}\, A_{nlm}\,  \vec \xi_{nlm} (R,\theta,\phi)\, {\rm  exp}[i \omega_{nlm}\, t] \, ,
\ee
\noindent
where the $\vec\xi_{nlm}$ are the eigenmodes.  (The $n=0$ modes are called f-modes, while the  $n \ge 1$ acoustic modes are called p-modes. 
Solar g-modes have not been definitively detected yet, and we will not consider them in this paper.)   For the most energetic modes, the surface velocity is dominated by its radial part, $v^r$, and for 
%
a spherically symmetric Sun, the radial components of the modes $\xi^r_{nlm}$ are 
$\propto Y_{lm}(\theta,\phi)$.

We will find it useful to separate the Sun's modes into two classes: $\sim 10^7$ resolvable modes, and then all the unresolvable ones. For our purposes, the unresolvable modes form a "confusion noise" background, in which the resolvable foreground modes are imbedded.   Besides unresolved oscillation modes, there are several noise sources that one might  call "seeing noise":  finite photon counts, Earth atmospheric effects, pixelization, etc.  (And there are {\it other} solar phenomena that are often not included in one's helioseismic model, such as 
sunspots.)  For the current study, it seems reasonable to  assume that surface motions from unresolved modes dominate the "noise".  And because the number of unresolved modes is vast, by the Central Limit Theorem that noise should be approximately Gaussian.   
\subsection{Coupling of nearly degenerate modes: a simple, two-mode illustration}
We emphasize that the main physical {\it reason} that the Sun's B-field can induce significant coupling is because modes that have the same $(n,l)$ values and nearby $m$ values are nearly degenerate in frequency. Therefore in this section we review some basic results from nearly-degenerate perturbation theory.
The effect is simplest to understand for non-dissipative systems.
For zero damping the perturbed fluid equations can be written in Hamiltonian form, so consider the following 2-D Hamiltionian: 
\begin{eqnarray}
\label{2d_examp1}
 \left( \begin{array}{cc}
1+ \delta &\ \   0 \\
 0 &\ \   1\ \end{array} \right) 
\end{eqnarray}
\noindent
(with $\delta$ real), whose eigenvectors are obviously $(1,0)$ and $(0,1)$.   Now we add a  small, off-diagonal perturbation:
\begin{eqnarray}
\label{2d_examp2}
 \left( \begin{array}{cc}
1+ \delta &\ \  \epsilon \\
 \epsilon^* &\ \  1\ 
\end{array} \right)
\end{eqnarray}
For simplicity, we will consider just the two limiting cases where  either $|\epsilon| << |\delta|$ or $|\epsilon| >> |\delta|$.
Case (i): $|\epsilon| << |\delta|$.  Then the two new eigenvalues are approximately $1 + \delta + |\epsilon|^2/\delta$ and 
$1 - |\epsilon|^2/\delta$ and the corresponding (unnormalized) eigenvectors are approximately $(1, \epsilon^*/\delta)$ and $(-\epsilon^*/\delta, 1)$, respectively.
E.g., if  $\epsilon \sim 10^{-4}$ and  $\delta \sim 10^{-3}$, then while the eigenvalues are perturbed by only
$\sim  |\epsilon|^2/\delta \sim 10^{-5}$, the components of the eigenvectors are shifted by $\sim  \epsilon/\delta \sim 10^{-1}$.
Case (ii): $|\epsilon| >> |\delta|$.  The two new eigenvalues are approximately $1 + \delta/2 + |\epsilon| $ and 
$1+ \delta/2 - |\epsilon| $ and the corresponding (unnormalized) eigenvectors are approximately $(\epsilon/|\epsilon| , 1 - 0.5\delta/|\epsilon| )$ and $(1 - 0.5 \delta/|\epsilon|, -\epsilon/|\epsilon|)$, resp.   That is, while the shift in the eigenvalues is only
$\emph{O}(\epsilon)$, the components of the eigenfunctions are shifted by $\emph{O}(1)$.

\subsection{B-field terms in the equations of motion, and their expansion into spherical harmonics }
Let $v^i$ be the perturbed (time-dependent) velocity field. Very schematically, the interior B-field adds an extra "forcing" term to the perturbed fluid equations of motion~\cite{Schnack2009}:
\be\label{forcing}
\rho \ddot  \xi^i = (non{\text{--}}B\ terms)\  + \gamma^i_j\,  \xi^j +   \kappa^i_{jk}\, \nabla^j \xi^k +  \eta^i_{jkl}  \nabla^j \nabla^k \xi^l
\ee
\noindent
where $\gamma^i_j$ is a sum of terms of the general form $\nabla B \, \nabla B$ (with two of the indices contracted; e.g., $\nabla_k  B^i \, \nabla_k  B_j$) and $B \nabla \nabla B$. 
 Similarly  $\kappa^i_{jk}$ is a sum of terms of the form $B \, \nabla B$  (e.g., $ B^i \, \nabla_k  B_j$)  and  $\eta^i_{jkl}$ is a sum of terms of the general form $e  B B$, where 
$e_{ij}$ is the spatial 3-metric. 
The tensors $\gamma^i_j$, $\kappa^i_{jk}$, and $\eta^i_{jkl}$ can all be expanded in tensor spherical
harmonics: basically tensorial versions of the $Y_{lm}$'s.  
Based on the preferred longitudes for sunspots, it seems a reasonable guess that the largest (non-axisymmetric) couplings will come from low-$m$ terms:  $m=\pm 1, \pm 2, \pm3, {\rm or} \pm 4$. 
%
%
%
Recall, from the standard rules of composition of angular momentum, that a perturbation $\propto Y_{lm}$ (with $m$ non-negative, say) can couple two modes $\propto Y_{l'm'}$ and $\propto Y_{l''m''}$, respectively,  only if (i) $|m'' - m'| = m$, (ii)$ | l' - l'' | < l < l' + l''$, and  (iii)$ l + l' + l''$ is even ( with the last condition essentially representing parity conservation).
The lowest order spherical harmonic connecting an $(n,l,m)$ and $(n,l,m+1)$ mode-pair  is a $Y_{21}$ term (since the coupling from a  $Y_{11}$ term must vanish by condition (iii) ), so it seems likely that this would usually be the dominant magnetic-perturbation term  
Similarly, it seems likely that the
dominant magnetic perturbation that connects modes $(n,l,m)$ and $(n,l,m+2)$  modes is a $Y_{22}$ term. 
Of course, the amplitude of B-field forcing terms will have some radial dependence, which should be reflected in the relative strengths of the mode couplings for different $n$ values.
%
\section{The magnitude of eigenmode mixing}
\label{Sec:mixing}
In this section we estimate the magnitude of B-field-induced eigenmode mixing in the Sun.  We do this in two steps: first, as a warm-up problem, in the limit of zero mode damping, and afterwards accounting for realistic damping rates.  Throughout this section we will analyze the coupling between mode pairs "as if no other modes existed", but, after some development, it will be easy to see that including all the modes in our analysis would have a negligible effect on any particular two-coupling.  The basic reason is that dimensionless mode-coupling parameters will typically have magnitude 
$\hat \lambda \sim 10^{-6}$ (or less), and the influence of any {\it other} modes on the coupling of some given mode-pair will be quadratic in these coupling parameters.

\subsection{Mode-mixing estimate for zero damping}
\label{Sec:zero-damp}
Observed p-modes typically have quality factors $Q \sim 10^3$, and therefore have damping times of order days.
This observation motivates making some initial, ballpark estimates in the limit of zero damping.  (We will find that our estimates based on this limit  are not at all accurate, but they will nevertheless provide an interesting basis for comparison.)
Here we will also make several other simplifying approximations.  First,  we will approximate the Sun as uniformly rotating at frequency
$\Omega$, so the Sun's "frozen in" $B$-field is time-independent in the Sun's rotating frame.  Therefore in this subsection we find it simplest to work in this rotating frame, since it allows us to use time-independent perturbation theory to estimate the amplitude of mode-mixing.
In an inertial rest frame, the frequency splitting of Solar modes with the same $(n,l)$ values, but different $m$ values, is dominated by the
kinematic term: $\Delta \nu \approx (\Omega/2\pi) \, \Delta m $.  However, in the rotating frame of the Sun (and ignoring for the moment the effect of the B-field), the mode-splitting is primarily due to the Coriolis effect: 
$\Delta \nu \approx c_{nlm} (\Omega/2\pi) \, \Delta m $, where the size of the coefficient $c_{nlm}$ is typically  $\sim 1 \%$~(\cite{Kosovichev96}).
Thus the $\delta$ term in our simple example,  Eqs.~(\ref{2d_examp1})--(\ref{2d_examp2}), would in this case be  $ \delta \sim |\Delta \nu/\nu| \sim  10^{-6}\,$ 
(for $|\Delta m| = 1$). Thus, for the idealized case of zero damping,  magnetic cross-coupling shift frequencies by only $\sim 10^{-6}$ (fractionally)  could still lead to 
${\cal O}(1)$ mixing between the eigenfunctions.    
%
%
%
\subsection{Mode-mixing estimate for realistic damping}
\label{Sec:mixing_with_damping}
We will now see that the mode-mixing estimates we made in the previous subsection are substantially altered when we incorporate realistic mode damping rates.  We will continue to approximate the Sun's rotation as uniform, but while in Sec.~\ref{Sec:zero-damp} we worked in the Sun's co-rotating frame, in this section we find it more convenient to work in the Sun's inertial frame.   Also in this subsection we will neglect spatial aliasing--effectively assuming that we are observing all $4\pi$ steradians of the Solar surface; we will analyze the effects of spatial aliasing in Sec.~\ref{sec:aliasing}. 

We return to our two-mode system, but this time described by the Langevin equation, which includes both damping and driving terms in addition to a mode-mode coupling term. For simplicity we will consider a pair of modes with quantum numbers  $(n,l,m)$ and $(n,l,m+1)$.  The extension to couplings between modes with $|\Delta m| = 2, 3, {\rm or} \, 4$ is trivial, and at the end of this subsection we will describe how to modify our equations to account for different values of $\Delta  m$.
 
Call the complex amplitudes of the two modes $a(t)$ and $b(t)$, respectively.  The Langevin equations for the coupled modes are then
\bea
\ddot a + \Gamma_a \dot a + \omega_a^2\, a & = & n_a(t) + \lambda e^{-i \Omega t} b(t)  \label{Langevin1}\\
\ddot b + \Gamma_b \dot b + \omega_b^2\, b  & = & n_b(t) + \lambda^* e^{i \Omega t} a(t)  \, . \label{Langevin2}
\eea

\noindent
Here $\omega_{a,b}$ are the two modes' oscillation frequencies,  $\Gamma_{a,b}$ are their damping rates, and the $n_{a,b}(t)$ are driving terms that have the statistical characteristics of noise.
%
 Since $\lambda$ is always small compared to $\omega_{a,b}^2$, we will treat it as a small perturbation. Fourier transforming the solutions to Eqs.~(\ref{Langevin1})-(\ref{Langevin2})  and expanding them through first order in $\lambda$, we find:
 
 
\bea
\tilde a(\omega) & = &  G_a(\omega)
\bigg[\tilde n_a(\omega) + \lambda  G_b(\omega+\Omega) \tilde n_b(\omega +\Omega)\bigg] \label{Langevin3}\\
\tilde b(\omega) & = & G_b(\omega)
\bigg[\tilde n_b(\omega) + \lambda^*  G_a(\omega - \Omega) \tilde n_a(\omega - \Omega)\bigg]  \, , \label{Langevin4}
\eea
%
 \noindent where the Green's functions $G_{a,b}(\omega)$ are explicitly given by
\bea
G_a(\omega) & \equiv &\big(-\omega^2 + i\omega \Gamma_a + \omega_a^2 \big)^{-1}  \label{Ga} \\
G_b(\omega) & \equiv& \big(-\omega^2 + i\omega \Gamma_b + \omega_b^2 \big)^{-1}  \, . \label{Gb}
\eea
and where our convention for Fourier transforms is
\be\label{def:FT}
 \tilde f(\omega) \equiv (2\pi)^{-1/2} \int_{-\infty}^{\infty} f(t) \, e^{-i \omega t} \, dt \, . 
 \ee  
Now, as a starting point for introducing our 
mode-mixing detection statistic, consider the following integral:
%
\be
\int{a^*(t)\, b(t) e^{-i \Omega t}  dt} \, ,
\ee
which by the convolution theorem equals
 \be\label{smokegunstatb0}
\int{\tilde a^*(\omega)\,  \tilde b(\omega + \Omega)  d\omega} \, .
\ee
To maximize the signal-to-noise, it will eventually prove useful to introduce an additional weighting factor $W(\omega)$ in the integrand of (\ref{smokegunstatb0}), so 
the modified version will be 
\be\label{smokegunstatb1}
\int{W(\omega)\, \tilde a^*(\omega)\,  \tilde b(\omega + \Omega)  d\omega} \, .
\ee
\noindent
But for now we will stick with the version in Eq.~(\ref{smokegunstatb0}).  We will also eventually want to restrict the limits of integration
in Eq.~(\ref{smokegunstatb1}), but for now we just leave them unspecified.
An expansion of Eq.~(\ref{smokegunstatb0}) through first order in $\lambda$ yields:
%
\bea
\int G^*_a(\omega) G_b(\omega + \Omega)  \bigg[\tilde n^*_a(\omega) \tilde n_b(\omega + \Omega)  \nonumber \\
+  \lambda^*  \bigg(|\tilde n_b (\omega + \Omega)|^2 G^*_b(\omega + \Omega) + |\tilde n_a (\omega)|^2 G_a(\omega) \bigg) \bigg] d\omega   \label{Gint_b}\, .
\eea
%
Next we will compare the relative sizes of the terms that are  linear in $\lambda^*$ (i.e., the B-dependent, "signal" terms") with the sizes of the $\lambda$-independent terms  (i.e., the B-independent, "background noise" terms).
It seems safe to approximate the noise as stationary, and to approximate its spectrum as flat over the very narrow region of interest.  Stationarity implies that noise amplitudes at different frequencies are uncorrelated.  Likewise
$|\tilde n_a(\omega)|^2$ and $|\tilde n_b(\omega + \Omega)|^2$  should be approximately equal, so we will refer to both as  
$|\tilde n_0|^2$.   
Similarly, $\Gamma_a$ and $\Gamma_b$  should be nearly the same, so for simplicity, in our estimates we will also take them to be equal: $\Gamma_a = \Gamma_b \equiv \Gamma$.  
Using the fact that $\tilde n_a(\omega)$ and 
$\tilde n_b(\omega + \Omega)$ are uncorrelated, the expectation value of (\ref{Gint_b}) becomes
\be
\hat \lambda^* \omega_0^2  |\tilde n_0|^2 \int{G^*_a(\omega) G_b(\omega + \Omega)\bigg(G_a(\omega) \, + \, G^*_b(\omega + \Omega)  \bigg)  \, d\omega  }\, .  \label{tripleG_2}
 \ee
where $\omega_0 \equiv (\omega_a + \omega_b)/2$,  $\hat \omega \equiv \omega - \omega_a$ and  $\Delta  \equiv (\omega_b - \omega_b)/2 = (\Omega + \epsilon)/2$ (so $\Delta$ is positive), and where we have defined $\hat\lambda \equiv \lambda/\omega_0^2$ (so $\hat\lambda$ is dimensionless).    Again, 
$\epsilon$ is typically $\sim 1\%$ the size of $\Omega$, so $\Delta$ is nearly equal to 
$\Omega$.
Then, neglecting terms that are cubic (or higher) in the small (compared to $\omega_a$) quantities $\hat\omega$, $\Omega$ and/or $\Gamma$, one easily shows that the 
sum $\big(G_a(\omega)\, + G^*_b(\omega + \Omega)  \big) $ in Eq.~(\ref{tripleG_2}) becomes
\be\label{gsum}
\frac{ -4\omega_a\hat\omega + i\Omega\,\Gamma } {\omega_a^2(4\hat\omega^2 + \Gamma^2)}
\ee
%
Similarly the term $\big[G^*_a(\omega) G_b(\omega + \Omega)\big]$  can be approximated as
\be\label{GaGb}
\bigg[ \omega_a^2\, \bigg(\Omega^2 + \Gamma^2 -4\Omega\hat\omega - 2i\, \Omega\,\Gamma \bigg)\bigg]^{-1}
\ee
 and so the expectation value of the "signal part" of our modified statistic  (\ref{smokegunstatb1}) becomes
%
\bea\label{sigint}
S = \int{W(\omega) 
\frac{\big(-4\hat\lambda^* \omega_0^2 \omega_a^{-3} |\tilde n_0|^2 \big) \big(\hat\omega - \frac{i}{4}  \Omega\Gamma \omega_a^{-1}\big)}{\big(4\hat\omega^2 + \Gamma^2\big)\big(\Omega^2 + \Gamma^2 -4\Omega\hat\omega - 2i\Omega\,\Gamma \big) }d\omega }\, .
\eea

We asserted at the beginning of this section that, while we have restricted attention to only two modes, including the effects of couplings to other modes would affect any 2-mode SNR only by higher order terms in the magnetic coupling parameters.   At this point, an easy way to see that is just
to add a third mode, with amplitude $c(t)$, to our  dynamical system system Eqs.~(\ref{Langevin1})--(\ref{Langevin2}), and repeat our calculations
down through Eq.~(\ref{Gint_b}).  We leave that as an exercise for the reader, after which the generalization to an arbitrary number of modes should  be obvious.


Again, we want to choose $W(\omega)$ to maximize the signal-to-noise, but before doing so, 
to simplify the analysis, we will make a couple more approximations.  First, because of the $\big(4\hat\omega^2 + \Gamma^2\big)$ factor in the denominator of the integrand in Eq.~(\ref{sigint}), the integral will be dominated by the region $\hat\omega \equiv (\omega - \omega_a) \alt {\rm few}\times \Gamma$.  In this region, the rms value of $\hat\omega$ is $\sim \Gamma$ (and below we will see that it is this rms value
that matters for the SNR) so in the factor  $\big(\hat\omega - \frac{i}{4}  \Omega\Gamma \omega_a^{-1}\big)$ the ratio of the second term to the first is typically $\sim \Omega/\omega_a \sim 10^{-4}$, so we will neglect that second term.   For the same reason, we can approximate
$\omega_0^2 \omega_a^{-3}$ by $\omega_0^{-1}$.

To further simplify the calculation, we will 
assume for the moment that one of the terms $\Gamma$ or $\Omega$ is significantly larger than the other; i.e., either $\Gamma >> \Omega$ or 
$\Omega >> \Gamma$.  In either case, in the term 
\be
\big(\Omega^2 + \Gamma^2 -4\Omega\hat\omega - 2i\, \Omega\,\Gamma \big) \, ,
\ee
the pieces $\big( -4\Omega\hat\omega - 2i\, \Omega\,\Gamma \big)$ can be neglected with respect to 
$\big(\Omega^2 + \Gamma^2\big) $.  
So Eq.~(\ref{sigint}) has now been approximated as
\bea\label{signalint}
S = \int{W(\omega) 
\frac{\big(-4\hat\lambda^* \omega_0^{-1} |\tilde n_0|^2 \big) \hat\omega }{\big(4\hat\omega^2 + \Gamma^2\big)\big(\Omega^2 + \Gamma^2  \big) }d\omega }\, .
\eea
To choose the optimal $W(\omega)$,  
we next need an expression for the
 "background noise" piece $N$, which arises from random correlations between the two modes.

We find $N$ easiest to estimate if we approximate the $\lambda$-independent part of the continuous integral (\ref{smokegunstatb1}) by 
the corresponding discrete sum over frequency bins, with bin width $\Delta \omega = 2\pi/T_{obs}$: 
\be\label{noise_est1}
N = \sum_i W(\omega_i)\, G^*_a(\omega_i) G_b(\omega_{i+h}) \tilde n^*_a(\omega_i) \tilde n_b(\omega_{i+h}) \frac{2\pi}{T_{obs}} \, .
\ee
where $h$ is the integer nearest to $\Omega/\Delta\omega$.
The terms $\tilde n^*_a(\omega_i)$ and $\tilde n_b(\omega_{i+h})$ are statistically independent, so the sum accumulates like a random walk.  Hence, using the same approximations as above, we have
\be\label{noise_est2}
\big< |N|^2 \big >= \sum_i |W(\omega_i)|^2\, \frac{\omega_0^{-4} |\tilde n_o|^4 \big(2\pi/T_{obs}\big)^2}{\big(\Omega^2 + \Gamma^2 \big)^2 }   \, 
\ee
%
Converting this discrete sum back to an integral, we
obtain
\be\label{noise_est3}
\big< |N|^2 \big >  = \int {W^2(\omega)\, \omega_0^{-4} \big(\Omega^2 + \Gamma^2\big)^{-2} |\tilde n_o|^4  \bigg(\frac{2\pi}{T_{obs}}\bigg)\, d\omega  }\, 
\ee
Optimizing $W(\omega)$ means maximizing $|S|$ 
for fixed $\big<N^2 \big >^{1/2}$.  Using the method of Lagrange multipliers, one finds that the
optimum choice is
\be\label{optW}
W(\omega) \propto  \frac{\hat\omega}{4\hat \omega^2 + \Gamma^2} \, 
\ee
\noindent
Thus the final form of our mode-mixing detection statistic (for $|\Delta m| = 1$) is 
\be\label{smokegunstatb2}
\int_{\omega_a - 5\Gamma}^{\omega_a + 5\Gamma}{\tilde a^*(\omega)\,  \tilde b(\omega + \Omega) \frac{\omega - \omega_a}{4(\omega - \omega_a)^2 + \Gamma^2} d\omega} \, .
\ee
The limits of integration here are somewhat arbitrary, but since the integrand
falls off rapidly for $|\omega - \omega_a| >> \Gamma$, the SNR should depend only weakly on the choice.  We should explain that the reason we restrict the integration range at all is just our intuition that restricting the region of integration limits the possibility of contamination from any artifacts in the spectra (e.g., from data gaps, any instrumental lines, etc.) .

 %
Plugging Eq.~(\ref{optW}) into Eq.~(\ref{signalint}) and using the approximation
\be\label{defint}
\int{\frac{\hat\omega^2}{(4\hat \omega^2 + \Gamma^2)^2} d\tilde\omega } \approx  \frac{\pi}{16\,  \Gamma} \, ,
\ee
\noindent
(the rhs of Eq.~(\ref{defint}) is actually the exact value of the integral when the limits of integration are taken to $\pm \infty$),
we find that $ |S| = (\pi/4) |\hat\lambda |  \omega_0^{-1} \Gamma^{-1}\big(\Omega^2 + \Gamma^2\big)^{-1} |\tilde n_o|^2 $ and that
\be
\big< |N|^2 \big >^{1/2}  = \frac{\pi}{\sqrt{8}}  \omega_0^{-2} \Gamma^{-1/2}\big(\Omega^2 + \Gamma^2\big)^{-1}  |\tilde n_o|^2 T_{obs}^{-1/2} \, .
\ee
Hence we arrive at
\be\label{snr1}
\frac{|S|}{\big< |N|^2 \big >^{1/2} }  = \frac{1}{\sqrt 2}\, |\hat\lambda|\, \bigg (\frac{\omega_0}{ \Gamma}\bigg)^{1/2} \,
\bigg(\omega_0 T_{obs} \bigg)^{1/2} \, 
\ee
so the SNR for a single mode-pair is 
\bea
SNR_{1-pair}\sim\frac{1}{\sqrt 2} \, |\hat \lambda| \, Q^{1/2}\bigg (\omega_0 T\bigg)^{1/2} \label{snr1} \\
\sim 0.18  \bigg(\frac{|\hat \lambda|}{10^{-5}}\bigg) \bigg(\frac{Q}{10^3}\bigg)^{1/2}  \bigg(\frac{\nu_0}{3.3 {\rm mHz}}\bigg)^{1/2}  \bigg(\frac{T_{obs}}{1\, {\rm yr}}\bigg)^{1/2} \label{2mode_est} 
\eea
\noindent  where $Q \equiv \omega_0/\Gamma$ and $\nu_0 \equiv \omega_0/(2\pi)$ 
Note that while we started with the simplifying assumption that either $\Gamma >> \Omega$ or 
$\Omega >> \Gamma$, or final result Eq.~(\ref{2mode_est}) is independent of which of these limits we are in.  It therefore seems reasonable to assume that Eq.~(\ref{2mode_est}) is also a fairly accurate estimate of the SNR for the intermediate case $\Gamma \approx \Omega$. 
Finally, while for simplicity of exposition, we so far have restricted to the  $\Delta m =1$ case, the generalization to $\Delta m = 2,3,4$ is trivial: in every numbered equation in this subsection, just replace $\Omega$ by
$\Delta m \, \Omega$.
\subsection{Summing the signal over mode pairs}
\label{sec:sum}
At first glance, the estimate (\ref{2mode_est}) does not seem very promising, unless the B-field is near the upper range of expectations.  
But Eq.~(\ref{2mode_est}) represents the SNR for just a single pair of modes, and it seems likely that the situation improves quite dramatically when one combines 
results from many mode pairs with the same $\Delta m$.  The reason is that the B-field coupling should be nearly phase-coherent over large numbers of mode pairs.
To be concrete, assume, e.g., that the dominant coupling for mode pairs with 
$\Delta m = 1$ comes from a  
$Y_2^{1}$ tensor perturbation.   This is a large-angular-scale perturbation, and so it seems likely that the phase of the perturbation is also coherent over a large 
range of radii.  (Of course, as the Sun rotates and drags the B-field with it, the complex amplitude of the perturbation rotates in the complex plane at the rate
$\Omega$.)
But then the signal part of our complex detection statistic should have (approximately) the same phase for some large number of mode pairs.  
%
%
Therefore if we add up the complex amplitudes of our detection statistic over mode pairs, the signal parts add coherently, and the total SNR (over all pairs) scales like $N_{pair}^{1/2}$, where $N_{pair}$ is the number of (phase-coherent) pairs. There are of order $10^7$ measured modes, and so about $10^7$ neighboring mode pairs, $(n,l,m)$ and $(n,l,m+1)$, and so the  SNR enhancement factor could be as high as $\sim 3000$. That's a rough upper limit on the enhancement factor.  And clearly, the extent to which this possible enhancement is realized depends in part on the radial scale over which the angular pattern of the B-field changes substantially.
But even plugging in $N=10^4$, with $Q = 3\times 10^3$ and $T_{obs} = 4\,$yr, one finds  B-field couplings with $|\hat \lambda|$ as small as  $2 \times 10^{-6}$ could be measured with $SNR \approx 12$- which is much more promising!  
Based on the fact that sunspots cluster at up to $4$ longitudes (at any time), it seems quite plausible that our method could yield information on spherical harmonics of the forcing tensor, Eq.~(\ref{forcing}), up to $|\Delta m| \approx 4$. 

%
%
%
%
%
\subsection{Robustness of our mode-mixing detection statistic against spatial aliasing}
\label{sec:aliasing}
In our formula for the SNR of our 2-mode detection statistic, Eq.~(\ref{2mode_est}), the amplitudes $a(t)$ and $b(t)$  are the true amplitudes of the modes corresponding to spherical harmonics $(l,m)$ and $(l,m')$, respectively.  But to date all helioseismology observations have been 
made either on the Earth or from satellites whose distance from the Earth is a very small fraction of $1\,$AU.   So with current telescopes we only 
have access to the half of the Sun facing us, and, when one accounts for the fact that we only measure the component of the Sun's surface velocity that is along the line of sight, one finds that our {\it effective} viewing area is closer to one-third of the Sun's surface.
This is the origin of spatial aliasing -- a mixing of the measured spherical harmonics.  Spatial aliasing is a rather large effect at any instant, so one might worry that it will swamp the mixing due to the Sun's interior B-field.  The aim of this subsection is to show that, in fact, spatial aliasing does {\it not} substantially degrade the power of our detection statistic.  First we 
will show analytically why this is the case.  We have also performed some simple simulations of the effects of aliasing, and we will show that our simulation results are consistent with our analytic estimates.
%
%
Let $(x,y,z)$ be inertial coordinates, with origin at the center of the Sun and $z$ along the Sun's spin axis, and then define
$(\theta, \phi)$ on the surface of the Sun in the usual way:  $cos\theta = z/r$, $sin\theta\, cos\phi = x/r$ , and $sin\theta\, sin\phi = y/r$, with $r = \sqrt{x^2 +y^2 + z^2}$. 
(Here and below we shall neglect the Sun's slight oblateness; i.e., we model its surface as a sphere.) 
Let $\hat n(t)$ be the unit vector from the center of the Sun to the observer (here assumed to be on or near the Earth), and let $\hat r$ be the unit radial vector from the Sun's origin to the location $(\theta, \phi)$.
On the Sun's surface, define $\sigma(t)$ by
\be
\sigma(\theta,\phi,t)  = {\rm max}\{0, \hat n\cdot \hat r\} \ ;
\ee
i.e., $\sigma(\theta,\phi,t)$ is $\hat n \cdot \hat r$ on the "front side" of the Sun (facing the observer) and zero on the "back side".  
%
\noindent
Define the time-dependent inner product between any two complex functions $f(\theta,\phi,t)$ and $g(\theta,\phi,t)$ on the Sun's surface by:
\be
\label{inner}
\left<  f \,|\, g \right> =  \int f^*g \, \sigma\,  d\Omega  \, .
\ee

Next we define the overlap function $\gamma_{l,m}^{m'}(t)$ by 
\be\label{def:gamma}
\gamma_{lm}^{m'}(t) = \big(N_{lm} N_{lm'}\big)^{-1/2}\left< Y_{lm}(\theta,\phi) \,|\,Y_{l\,m'}(\theta,\phi) \right>  \, ,
\ee
where $\big(N_{lm} N_{lm'}\big)^{-1/2}$ is just a time-averaged normalization factor, specifically:
\be\label{def:Nlm}
N_{lm} \equiv \frac{1}{yr} \int_0^{1yr} \left< Y_{lm} \,|\,Y_{l\,m} \right> \,dt \, .
\ee
The overlap function $\gamma_{lm}^{m'}(t)$ varies on the timescale of a year because that is the timescale
on which the  "visible half" of the Sun (from the Earth) varies.  Also, clearly, 
$ \gamma_{lm}^{m'}(t) = \gamma_{lm'}^{m\, *}(t)$.

Now, continuing with our 2-mode example, imagine for simplicity that the Sun's perturbed radial velocity $\delta v^r$  at the surface $R$
 is the sum of only two modes:
\be\label{dvr}
\delta v^r (\theta, \phi) = a(t)\, Y_{l m}(\theta, \phi)   + b(t)\, Y_{l m'}(\theta, \phi) \, . 
\ee
\noindent
where the amplitudes (but not the phases) of $a(t)$ and $b(t)$ are slowly varying (due to damping and excitation) and where, again, $\omega_b - \omega_a
\approx (m' - m)\,\Omega$.  Now let us define the observed values of these amplitudes, $a_{ob}(t)$ and  $b_{ob}(t)$ by
\be\label{aob0}
a_{ob}(t) \equiv \left<  \delta v^r \,|\, Y_{lm}\right>  \, \ \ \  b_{ob}(t) \equiv \left<  \delta v^r \,|\, Y_{lm'}\right>  
\ee
so that 
\bea
a_{ob}(t) &=&   \gamma_{lm}^{m}(t)\, a(t) + \gamma_{lm}^{m'}(t)\, b(t) \label{aob1}\\
b_{ob}(t)  &=&   \gamma_{lm'}^{m'}(t)\, b(t) + {\gamma_{lm'}^{m}}\, a(t)  \, . \label{aob2}
\eea
Transforming to the Fourier domain, we then have
\bea
\tilde a_{ob}(\omega) &=& \int \!\!\bigg[\tilde\gamma_{lm}^{m}(\omega \!- \!\!\omega^{\prime}) \, \tilde a(\omega') + \tilde\gamma_{lm}^{m'}(\omega \!-\!\! \omega^{\prime})\, \tilde b(\omega')\bigg]\!d\omega^{\prime}  \label{observed_amps1}\\  
\tilde b_{ob}(\omega) &=& \int \!\!\bigg[\tilde\gamma_{lm'}^{m'}(\omega \!- \!\!\omega^{\prime})\, \tilde b(\omega') +  \tilde\gamma_{lm^{\prime}}^m(\omega\! -\!\! \omega^{\prime})\, \tilde a(\omega')\bigg]\!d\omega^{\prime}  \, . \label{observed_amps2}
\eea

Now let us see what happens if in Eq.(\ref{smokegunstatb1}) we replace $\tilde a(\omega)$ and $\tilde b(\omega)$ by the corresponding 'observed' amplitudes, given by Eqs.~(\ref{observed_amps1})-(\ref{observed_amps2}).  The "background noise" piece 
\be\label{bkgd1}
 \int W(\omega) G^*_a(\omega) G_b(\omega + \Delta m\,\Omega) \tilde n^*_a(\omega) \tilde n_b(\omega + \Delta m\,\Omega)
\ee
gets augmented by one cross-term that is quadratic in $\tilde n_a$:
\bea
\int \int \int W(\omega) G^*_a(\omega') G_a(\omega'' + \Delta m\,\Omega)  \tilde n^*_a(\omega')\nonumber \\
 \tilde n_a(\omega'' + \Delta m\,\Omega) \tilde\gamma_{lm}^{m\,*}(\omega-\omega')\tilde\gamma_{lm'}^{m}(\omega + \Delta m\,\Omega -\omega'') d\omega \ d\omega'  d\omega''    \label{bkgd2}
\eea
as well as a similar term that is quadratic in $\tilde n_b$.
The convolution factors $\tilde\gamma_{lm}^{m\,*}(\omega-\omega')$ and $\tilde\gamma_{lm'}^{m}(\omega-\omega'')$
effectively "smear" $\tilde n^*_a(\omega) $ and $\tilde n^*_a(\omega + \Delta m\,\Omega) $ over frequency bands of width $\sim 2\pi/{\rm yr}$.
However the key point is that $\Omega >> 2\pi/{\rm yr}$, so even these smeared bands are non-overlapping, and hence their values are {\it not} correlated -- unlike the case the for signal terms.  It is easy to see that the same is true for the aliasing terms quadratic in 
$\tilde n_b$.  So aliasing modifies the background noise piece by only a small fraction of its value.
To illustrate how this works, Fig.~1 simulates the build-up over frequency of three pieces of the our detection statistic: the 
$\lambda$-independent noise piece given by (the continuous version of) Eq.(\ref{noise_est1}), the $\lambda$-dependent signal piece given by Eq.(\ref{signalint}), and the aliasing contribution given by the sum of Eq.(\ref{bkgd2}) and the corresponding term that is quadratic in $\tilde n_b$.  The specific parameters chosen for his particular simulation were $l=10, m = 4, m' = 5$,  $T_{obs} = 1/$yr, $\omega_0 = 0.02 s^{-1}$,
$\Omega = 3\times 10^{-6} {s}^{-1}$, $\Gamma = 10^{-3} \omega_0$, and $\omega_{a,b} = \omega_0 \pm \Omega/2$.  We also take $n_a$ and 
$n_b$ to be uncorrelated white noises with the same amplitude, which should be a good approximation over the narrow frequency band of interest.  
For the sake of visual clarity, we took 
$\hat\lambda = 3 \times 10^{-4}$, with is probably unphysically large, but does not affect the relative sizes of the "unaliased" background noise piece and the aliasing piece.   The take-home point of Fig.~1  is that the
aliasing contribution is only a modest fraction of the full noise, and so has little effect on our SNR estimates.
Fig.~1 displays just one realization of $n_a$ and $n_b$, but is a typical result. 
%
\begin{figure}
\vskip -1in
\label{snr_buildup}
 \centerline{\includegraphics[width=\columnwidth]{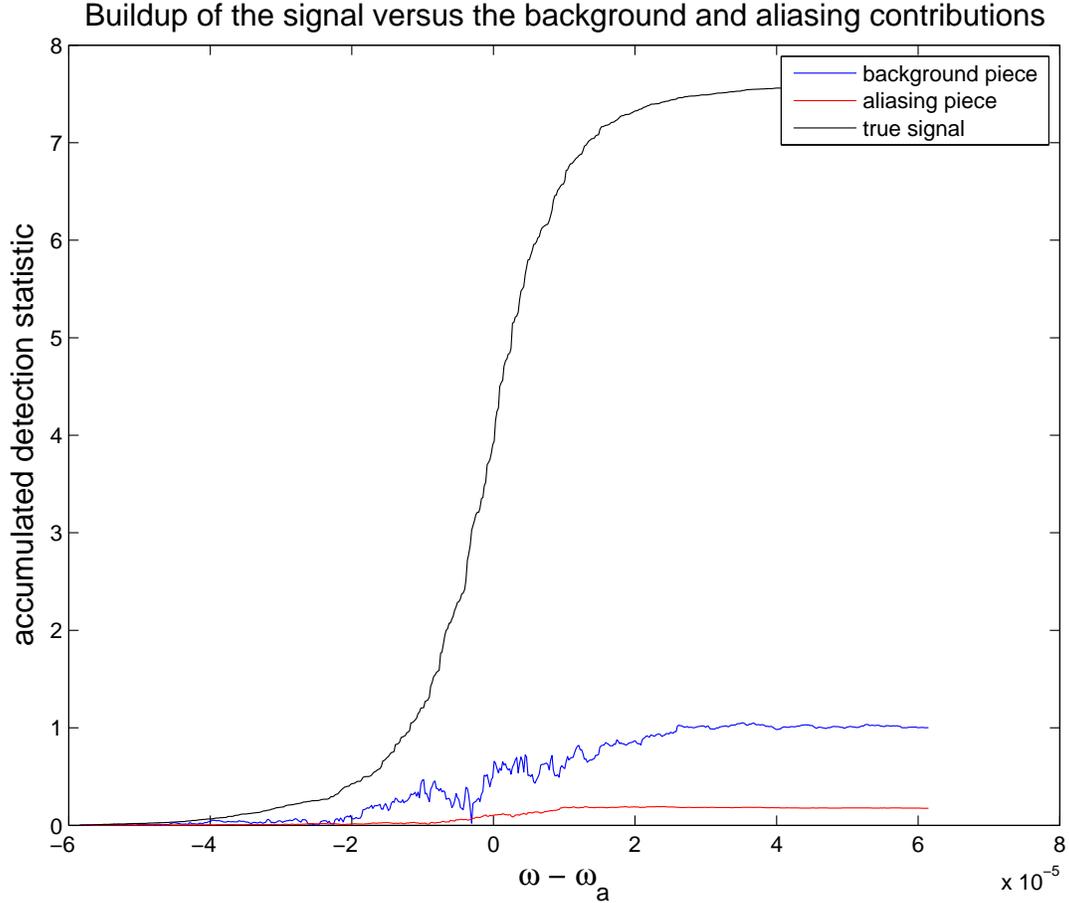}}
 \vskip -1in
\caption{
Illustrates the build-up, over the frequency integration band, of three different contributions to our mode-mixing detection statistic:  the random noise piece, the "true signal" piece, and a contribution from spatial aliasing/leakage. 
The specific parameters for this particular case were $l=10, m = 4, m' = 5$, $T_{obs} = 1\,$yr, and $\lambda = 3\times 10^{-4}$.  
The main point is that the aliasing contribution is much smaller than the random noise contribution, and so has only a very modest impact on the SNR of our detection statistic.}
\end{figure}
\noindent
To summarize, if one Fourier transforms the observed mode amplitudes, aliasing adds sidebands to the mode "lines".  But for integration times of order a year or more, the separation of these sidebands from the carrier is much less than the separation between the mode lines, and so aliasing does not induce the mode correlations that our detection statistic searches for.   We suggest that this method of mitigating the effects of spatial aliasing might prove useful in other sorts of helioseismological studies as well.
\section{Summary, caveats and future work}
\label{Sec:summary}
In this paper we investigated the possibility of using mode-mixing to probe the Sun's interior B-field. 
%
We concentrated on modes with the same $(n,l)$ values and nearby $m$ values, since such modes are nearly degenerate in frequency, which enhances mode-mixing.   
We constructed a novel mode-mixing detection statistic for this effect, Eq.~(\ref{smokegunstatb2}), and we showed that for long observation times (of order a year or more), our statistic is quite robust against the effects of spatial aliasing.   We estimated the SNR for our detection statistic, for realistic mode damping.  The detectability of mode-mixing is enhanced by a couple of factors, in addition to the near-degeneracies.  First, the SNR grows like $N_{cyc}^{1/2}$, where $N_c = \nu_0 T_{obs}$ is the number of observed oscillation cycles.  For a year's worth of observation of five-minute oscillations,  $N_{cyc}^{1/2} \sim 300$.  
Second, we argued that the phase of the mode-mixing is likely approximately constant over a large range of $(n,l)$ values, for fixed $\Delta m$.
So by adding up the complex values of our detection statistic Eq.~(\ref{smokegunstatb2}) over $N_{pair}$ mode pairs, the total SNR  
should grow roughly as $N_{pair}^{1/2}$, where $N_{pair}$ is the number of mode pairs with similar mixing angle. 
$N_{pair}$ could be as large as $\sim 10^7$ , 
but even assuming $N_{pair} \sim 10^4-10^5$, Eq.~(\ref{2mode_est}) suggests that  couplings as small as $|\hat \lambda| \agt 10^{-6}$ should be 
detectable.   

As caveats, we here remind the reader of some effects that we have {\it not} yet taken into account in our analyses.  First, for simplicity, all our analyses took the Sun to be uniformly rotating.   It would be more accurate to describe the Sun's angular velocity as a constant $\Omega_0$
(some weighted average of the angular velocity field) plus an axisymmetric perturbation $\delta\Omega(r, \theta)$.  However we would argue that this improvement would not substantially affect our estimate of the mode-mixing SNR.  The reason is that, as we have seen, the physical mechanism that is most important for causing mode-mixing to saturate is mode damping, and typically $\Gamma > \delta\Omega$.  
 I.e., on the damping timescale $\Gamma^{-1}$, the non-uniform part of the angular velocity, $\delta\Omega$, is too small to cause much "re-arrangement" of fluid and magnetic field inside the Sun.   But we have not actually demonstrated this, so include that as a caveat. 
 

Secondly, and perhaps most importantly, we have not yet tried to assess the likely impact of systematic errors on measurements of mode mixing.
There are quite a few known instrumental effects, such as pixelization, whose impact we could reasonably try to assess.  However,  inversions of helioseismology data today also reveal effects that are clearly spurious but of unknown origin, such as the infamous "center-to-limb" 
effect, see ~\cite{duvall09,baldner12,zhao13}.  It is hard to assess the impact of systematics that are not understood, which is a problem that our proposed method shares with much of the rest of helioseismology.
%
%

Regarding future work, Cutler \& Woodard have recently begun to calculate the summed version of our mode-mixing detection statistic using SDO/HMI data.  
\section*{Acknowledgments}
This work was carried out at the Jet Propulsion Laboratory, California Institute of Technology, under contract to the National Aeronautics and Space Administration.   Special thanks go to Martin Woodard for a great many helpful and informative discussions, general encouragement,  and for carefully reading and improving a draft of this manuscript.  I also owe special thanks to Marco Velli and Neil Murphy for getting me involved in this subject and for many useful discussions and overall encouragement.  Also, Stuart Jeffries, Douglas Gough, Charles Baldner and Tim Larson were all very generous with their time in helping educate me about this subject.
%
\bibliographystyle{aasjournal}

\bibliography{new_merged}

\begin{thebibliography}{}
\expandafter\ifx\csname natexlab\endcsname\relax\def\natexlab#1{#1}\fi
\providecommand{\url}[1]{\href{#1}{#1}}

\bibitem[{{Antia} {et~al.}(2013){Antia}, {Chitre}, \& {Gough}}]{antia2013}
{Antia}, H.~M., {Chitre}, S.~M., \& {Gough}, D.~O. 2013, \mnras, 428, 470

\bibitem[{{Baldner} \& {Schou}(2012)}]{baldner12}
{Baldner}, C.~S., \& {Schou}, J. 2012, \apjl, 760, L1

\bibitem[{{Duvall} \& {Hanasoge}(2009)}]{duvall09}
{Duvall}, Jr., T.~L., \& {Hanasoge}, S.~M. 2009, in Astronomical Society of the
  Pacific Conference Series, Vol. 416, Solar-Stellar Dynamos as Revealed by
  Helio- and Asteroseismology: GONG 2008/SOHO 21, ed. M.~{Dikpati},
  T.~{Arentoft}, I.~{Gonz{\'a}lez Hern{\'a}ndez}, C.~{Lindsey}, \& F.~{Hill},
  103

\bibitem[{{Fuller} {et~al.}(2015){Fuller}, {Cantiello}, {Stello}, {Garcia}, \&
  {Bildsten}}]{fuller_etal_2015}
{Fuller}, J., {Cantiello}, M., {Stello}, D., {Garcia}, R.~A., \& {Bildsten}, L.
  2015, Science, 350, 423

\bibitem[{{Gough} \& {Taylor}(1984)}]{gough_taylor84}
{Gough}, D.~O., \& {Taylor}, P.~P. 1984, \memsai, 55, 215

\bibitem[{{Gough} \& {Thompson}(1990)}]{gough1990}
{Gough}, D.~O., \& {Thompson}, M.~J. 1990, \mnras, 242, 25

\bibitem[{{Kosovichev}(1996)}]{Kosovichev96}
{Kosovichev}, A.~G. 1996, \apjl, 469, L61

\bibitem[{{Miesch}(2005)}]{miesch05}
{Miesch}, M.~S. 2005, Living Reviews in Solar Physics, 2, 1

\bibitem[{{Schad} {et~al.}(2011{\natexlab{a}}){Schad}, {Roth}, \&
  {Timmer}}]{schad1}
{Schad}, A., {Roth}, M., \& {Timmer}, J. 2011{\natexlab{a}}, Journal of Physics
  Conference Series, 271, 012079

\bibitem[{{Schad} {et~al.}(2011{\natexlab{b}}){Schad}, {Timmer}, \&
  {Roth}}]{schad2}
{Schad}, A., {Timmer}, J., \& {Roth}, M. 2011{\natexlab{b}}, \apj, 734, 97

\bibitem[{{Schad} {et~al.}(2012){Schad}, {Timmer}, \& {Roth}}]{schad3}
---. 2012, Astronomische Nachrichten, 333, 991

\bibitem[{{Schad} {et~al.}(2013){Schad}, {Timmer}, \& {Roth}}]{schad4}
---. 2013, \apjl, 778, L38

\bibitem[{{Schnack}(2009)}]{Schnack2009}
{Schnack}, D.~D., ed. 2009, Lecture Notes in Physics, Berlin Springer Verlag,
  Vol. 780, {Lectures in Magnetohydrodynamics}

\bibitem[{{Stello} {et~al.}(2016){Stello}, {Cantiello}, {Fuller}, {Huber},
  {Garc{\'{\i}}a}, {Bedding}, {Bildsten}, \& {Aguirre}}]{stello_etal_2016}
{Stello}, D., {Cantiello}, M., {Fuller}, J., {et~al.} 2016, \nat, 529, 364

\bibitem[{{Woodard} {et~al.}(2012){Woodard}, {Schou}, {Birch}, \&
  {Larson}}]{woodard12}
{Woodard}, M., {Schou}, J., {Birch}, A.~C., \& {Larson}, T.~P. 2012, \solphys,
  179

\bibitem[{{Woodard} {et~al.}(2013){Woodard}, {Schou}, {Birch}, \&
  {Larson}}]{woodard13}
---. 2013, \solphys, 287, 129

\bibitem[{{Zhao} {et~al.}(2013){Zhao}, {Bogart}, {Kosovichev}, {Duvall}, \&
  {Hartlep}}]{zhao13}
{Zhao}, J., {Bogart}, R.~S., {Kosovichev}, A.~G., {Duvall}, Jr., T.~L., \&
  {Hartlep}, T. 2013, \apjl, 774, L29

\end{thebibliography}
\end{document}